\documentclass[%
 aps,
 reprint,
 superscriptaddress,
 amsmath,amssymb,
]{revtex4-2}
   
\usepackage{graphicx}
\usepackage{dcolumn}
\usepackage{bm}
\usepackage{float}
\usepackage{subfiles}
\usepackage{changes}
\usepackage{gensymb}
\usepackage{bm}
\usepackage{soul}

\usepackage[super]{nth}
\usepackage[normalem]{ulem}
\usepackage[
bookmarks=true,
colorlinks=true,
linkcolor=magenta,
urlcolor=blue,
citecolor=blue,
pdftex,
bookmarks=true,
linktocpage=true, 
hyperindex=true
]{hyperref}
\usepackage{hyperref}
\usepackage{siunitx}
\usepackage[super]{nth}
\usepackage{standalone}

\begin{document}

\title{Universal zero-crosstalk photonic integration via slab-engineered mode hybridization}
\newcommand{\KAIST}{School of Electrical Engineering, Korea Advanced Institute of Science and Technology, Daejeon 34141, Republic of Korea}
\newcommand{\KIST}{Center for Quantum Technology, Korea Institute of Science and Technology, Seoul 02792, Republic of Korea}
\newcommand{\KHU}{Department of Physics, Kyung Hee University, Seoul 02447, Republic of Korea}

\author{Kyungtae Kim}
\author{Yoseph Shin}
\author{Seungyong Lee} 
\author{Inki Kim}  
\affiliation{\KAIST}

\author{Hyeyoon Jeon} 
\affiliation{\KIST}
\affiliation{\KHU}

\author{Jibaek Song} 
\author{Minseop Lee}
\author{Sanghyeon Kim} 
\affiliation{\KAIST}

\author{Hyounghan Kwon} 
\affiliation{\KIST}

\author{Hojoong Jung} 
\affiliation{\KIST}
\affiliation{\KHU}

\author{Sangsik Kim}
\email{sangsik.kim@kaist.ac.kr}
\affiliation{\KAIST}

\begin{abstract}
Photonic integrated circuits have emerged as a scalable platform for optical computing, communication, and quantum technologies, where high-fidelity optical processing is essential. 
However, as photonic systems scale in complexity, inter-channel crosstalk accumulates across cascaded components, fundamentally degrading signal fidelity, limiting system-level performance, and constraining integration density.
Existing crosstalk-suppression strategies rely on specialized nanostructures or platform-specific designs, hindering their adoption in standard foundry processes and across diverse material systems.
Here we establish a universal and foundry-compatible route to eliminating crosstalk based on slab-engineered mode hybridization in standard rib waveguides. 
By tailoring the slab thickness, mode hybridization induces anisotropic modal perturbations that enable complete cancellation of coupling between adjacent waveguides.
We experimentally demonstrate zero-crosstalk across diverse material platforms, including silicon-on-insulator, silicon nitride, thin-film lithium niobate, and germanium-on-insulator, spanning wavelengths from the visible to the mid-infrared.
Our approach provides a manufacturable route toward scalable, high-fidelity, and high-density photonic integration, overcoming the long-standing trade-off between signal fidelity and integration density in large-scale photonic systems.
\end{abstract}
\maketitle

Photonic integrated circuits (PICs) are increasingly central to high-bandwidth interconnects \cite{Popovic2015single, Kippenberg2017microresonator, Popovic2018integrating, Bergman2018recent, Loncar_Wang2018integrated, Tsang2024silicon, Lethold014coherent, Lethold2023resonant}, energy-efficient AI computation \cite{Bogaerts2020programmable, Shen2017deep, Lin2018all,  Kippenberg2021parallel, Shastri2021photonics, Loncar2022integrated, Wang2024integrated,Tsang2024ultrafast, Liang2024highspeed}, and scalable quantum processing \cite{Daoxin2023verylarge, Qiang2018large,Wang2020integrated, Elshaari2020hybrid, Arrazola2021quantum,Kim2025integrated, KiYoul2022quantum, Poon2022monolithically}, enabling large-scale optical systems where signal fidelity and integration density become critical.
As photonic systems grow in complexity, their performance is no longer limited by individual devices but by the signal integrity across densely integrated components. 
In particular, inter-channel crosstalk---driven by evanescent coupling between proximate waveguides---grows rapidly with integration density.
Although often negligible at the individual component level, such parasitic crosstalk accumulates across cascaded stages, progressively degrading signal fidelity and limiting system-level performance \cite{Shekhar2024roadmapping, Daoxin2012passive}.
The conventional route---increasing the inter-waveguide spacing---suppresses crosstalk only at the expense of integration density and chip footprint, having a fundamental trade-off between signal integrity and scalability.

Extensive efforts have been devoted to mitigating optical crosstalk, including plasmonics \cite{Oulton2008hybrid,Haffner2015all,Kim2015mode}, inverse design \cite{Shen2016increasing}, waveguide superlattices \cite{Song2015high,Gatdula2019guiding}, and artificial gauge-field approaches \cite{Lumer2019light,Zhou2023agf}.
While these methods can reduce crosstalk under specific conditions, they typically rely on intricate geometries, subwavelength features, or platform-specific designs, limiting compatibility with standard foundry processes and hindering scalability across material systems. 
Subwavelength-grating metamaterials have also been explored as an alternative approach for controlling inter-waveguide coupling \cite{Jahani2018controlling,Mia2020exceptional,Kabir2023anisotropic}, enabling compact and high-performance devices \cite{Shin2025anisotropic,Ahmed2021ultra, Ahmed2023high,Mia2023broadband}. 
However, their reliance on deeply subwavelength patterning typically requires electron-beam lithography and fully etchable materials, posing significant challenges for large-scale manufacturing and restricting applicable material platforms.
As a result, a general and manufacturable route to eliminating crosstalk, compatible with standard foundry processes and applicable across material platforms and wavelength regimes, remains an open challenge.

\begin{figure*}[!htbp]
    \includegraphics[width=0.95\linewidth]{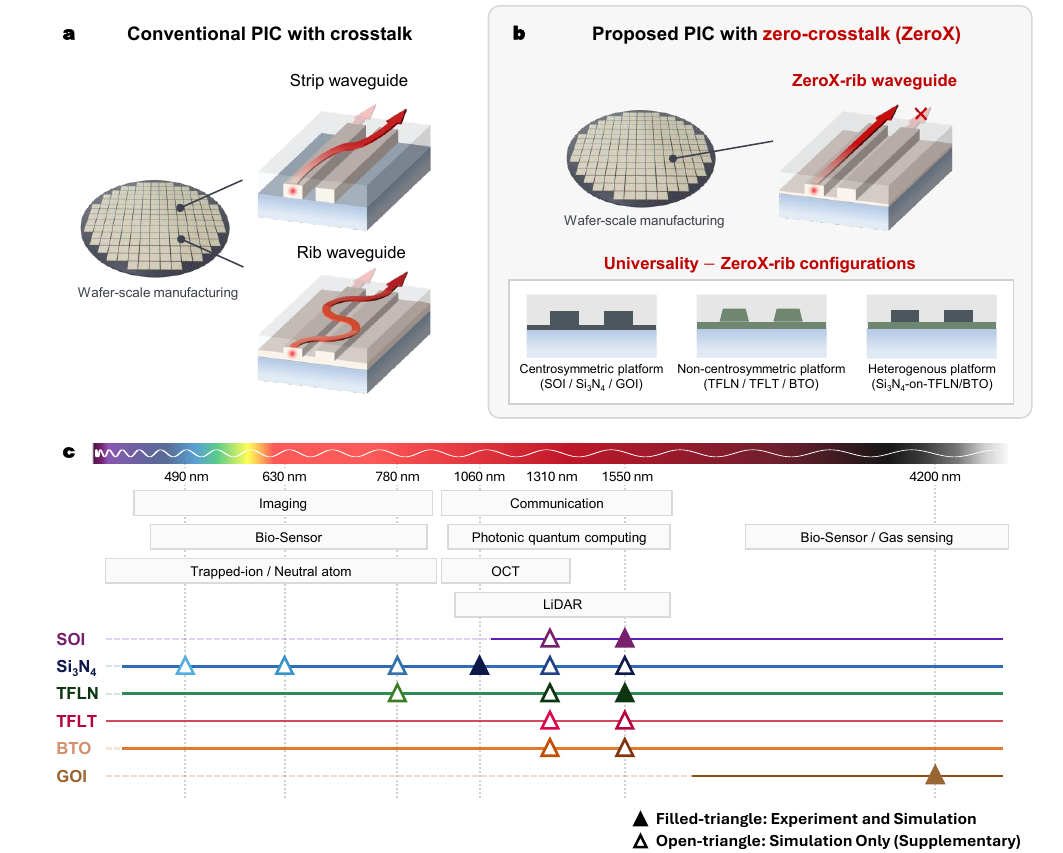}
    \caption{\textbf{Universal zero-crosstalk (zeroX) rib waveguides for scalable photonic integration.}
    \textbf{a,} Conventional strip and rib waveguides exhibit inter-waveguide crosstalk that limits fidelity and integration density of photonic integrated circuits (PICs).
    \textbf{b,} Proposed slab-engineered zeroX-rib waveguide that achieves complete crosstalk suppression. 
    The zeroX-rib mechanism is foundry-compatible and broadly applicable across centrosymmetric (e.g., SOI, Si$_3$N$_4$, GOI), non-centrosymmetric (e.g., TFLN, TFLT, BTO), and heterogeneous (e.g., Si$_3$N$_4$-on-TFLN/BTO) photonic platforms. 
    \textbf{c,} Spectral transparency windows of representative PIC material platforms and their potential applications, spanning the visible to the mid-infrared.
    Filled-triangle markers denote experimentally validated zeroX-rib demonstrations in this work (Fig.~\ref{Fig4}); 
    open-triangles indicate simulation results (Supplementary Section~1).
    SOI, silicon-on-insulator; $\mathrm{Si_3N_4}$, silicon nitride; TFLN, thin-film lithium niobate; TFLT, thin-film lithium tantalate; BTO, barium titanate; GOI, germanium-on-insulator.
   }
	\label{Fig1}
\end{figure*}

Here we establish a universal route to eliminating crosstalk based on slab-engineered mode hybridization in standard rib waveguides, which is platform-independent, compatible with standard foundry processes, and applicable across diverse wavelength regimes. 
We show that slab thickness serves as a previously underexplored degree of freedom that governs hybridization between $\text{TM}_0$ and $\text{TE}_1$ modes, enabling anisotropic modal perturbations that completely cancel inter-waveguide coupling.
Since this mechanism originates from intrinsic modal interactions inherent to rib geometries, the resulting zero-crosstalk behavior can be achieved across material platforms and spectral regimes.
Figure~\ref{Fig1} illustrates the proposed zero-crosstalk (zeroX) rib concept and its universality across diverse photonic platforms.
We experimentally validate this universality by demonstrating zeroX-rib waveguides in silicon-on-insulator (SOI), silicon nitride ($\mathrm{Si_3N_4}$), thin-film lithium niobate (TFLN), and germanium-on-insulator (GOI) platforms spanning wavelengths from the visible to the mid-infrared.
Together, these results establish a manufacturable pathway toward scalable, high-fidelity, and high-density photonic integration.\\

\begin{figure*}[!htbp]
    \includegraphics[width=0.85\linewidth]{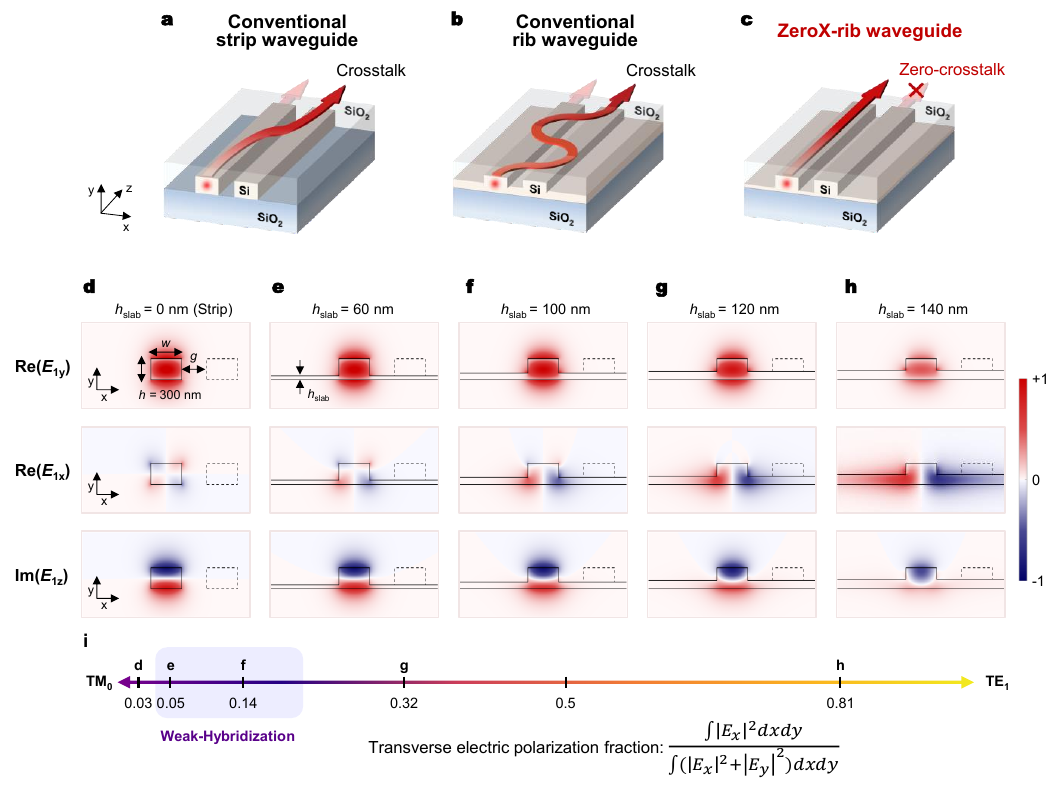}
    \caption{\textbf{Slab-engineered mode hybridization for zero-crosstalk operation.}
    \textbf{a--c,} Coupled waveguide configurations: 
    \textbf{a,} strip waveguide with crosstalk, 
    \textbf{b,} conventional rib waveguide with strong crosstalk due to the slab, and
    \textbf{c,} slab-engineered zeroX-rib waveguide with complete crosstalk cancellation at the same pitch.  
    \textbf{d--h,} Normalized electric-field components of the TM$_0$ mode of an isolated waveguide for slab thickness $h_{\text{slab}}=0$, 60, 100, 120, and 140~nm on an SOI platform with a silicon (Si) core and an oxide ($\mathrm{SiO_2}$) cladding. 
    Solid lines indicate the isolated waveguide; dashed regions indicate the perturbation region of the adjacent waveguide that governs coupling.
    The evolution reveals anisotropic redistribution of $\operatorname{Re}(E_{1x})$ and $\operatorname{Im}(E_{1z})$ with increasing slab thickness,
    while $\operatorname{Re}(E_{1y})$ remains largely unchanged.
    \textbf{i,} TE polarization fraction of the TM$_0$ as a function of $h_{\text{slab}}$, showing progressive hybridization between TM$_0$ and TE$_1$.
    The purple-shaded region marks the weakly hybridized regime, where crosstalk can be effectively canceled.
    Other geometric parameters: height $h=300~\text{nm}$, core width $w=550~\text{nm}$, and gap $g=450~\text{nm}$, and the wavelength is $\lambda_0=1550~\text{nm}$. 
    }
	\label{Fig2}
\end{figure*} 

\noindent \textbf{\large Results}\\
\noindent \textbf{Slab engineering for mode hybridization}\\
We begin by revisiting inter-waveguide coupling in conventional strip and rib waveguides.
In strip waveguides, strong evanescent overlap leads to pronounced crosstalk at reduced spacing, setting a fundamental limitation on integration density.
Rib waveguides, despite their additional geometric degree of freedom, typically exhibit similar or even stronger coupling due to the slab, which reduces modal confinement and introduces additional radiative coupling pathways.
However, this conventional behavior breaks down at a specific slab thickness, where a zero-crosstalk condition emerges at fixed waveguide geometries.
Figures~\ref{Fig2}a--c show this transition across three representative regimes: 
conventional strip and rib waveguides, both exhibiting pronounced crosstalk, and a slab-engineered zeroX-rib waveguide that achieves complete crosstalk suppression at the same pitch.

To elucidate the physical origin of this transition, we analyze the evolution of modal field profiles as the slab thickness $h_{\text{slab}}$ increases on an SOI platform.
Figures~\ref{Fig2}d--h show the normalized electric-field components of the uncoupled (isolated waveguide) $\mathrm{TM}_0$ mode, where the solid boundary denotes the isolated waveguide and the dashed region indicates the perturbation region of the adjacent waveguide that governs coupling. 
For the strip configuration ($h{_\text{slab}}=0$), the modal field distribution remains largely symmetric due to the structural symmetry.
Introducing a slab layer breaks this symmetry and induces anisotropic redistribution of the modal fields.
While $\operatorname{Re}(E_{1y})$ remains relatively unchanged, the bipolar (positive-negative) distributions of $\operatorname{Re}(E_{1x})$ and $\operatorname{Im}(E_{1z})$ undergo significant transformations as $h_{\text{slab}}$ increases.
In particular, $\operatorname{Re}(E_{1x})$ evolves from a vertically balanced bipolar distribution toward a laterally dominant profile, resulting in a predominantly negative field in the perturbation region. 
Similarly, $\operatorname{Im}(E_{1z})$ undergoes a vertical shift, leading to a predominantly positive field distribution within the same region. 
Because inter-waveguide coupling is governed by the modal field distribution within this perturbation region, this anisotropic and sign-asymmetric field reconfiguration directly modifies the effective coupling between adjacent waveguides.

The observed field reconfiguration originates from hybridization between the $\text{TM}_0$ and $\text{TE}_1$ modes.
To quantify this effect, we evaluate the transverse-electric (TE) polarization fraction, where a value near 0.5 indicates maximal hybridization (see Methods). 
As shown in Fig.~\ref{Fig2}i, the TE polarization fraction of the $\text{TM}_0$ mode increases with $h_{\text{slab}}$, indicating progressively stronger mode mixing induced by the slab. 
This hybridization enhances the $E_x$ components of the nominal TM$_0$ mode, resulting in a weakly hybridized quasi-TM$_0$ state.
Importantly, zero-crosstalk emerges only within this weakly hybridized regime (highlighted in purple), where the relative contributions of different field components are balanced to cancel inter-waveguide coupling. 
Beyond a critical slab thickness, the mode transitions toward a TE$_1$-like state, and the coupling reappears.
Although other geometric parameters (e.g., core width) can influence hybridization, the slab thickness primarily governs the modal field reconfiguration that determines coupling.
Slab engineering thus provides a direct route to achieving zero-crosstalk through controlled mode hybridization (see Supplementary Section~2).

\begin{figure*}[!htbp]
	\includegraphics[width=0.85\linewidth]{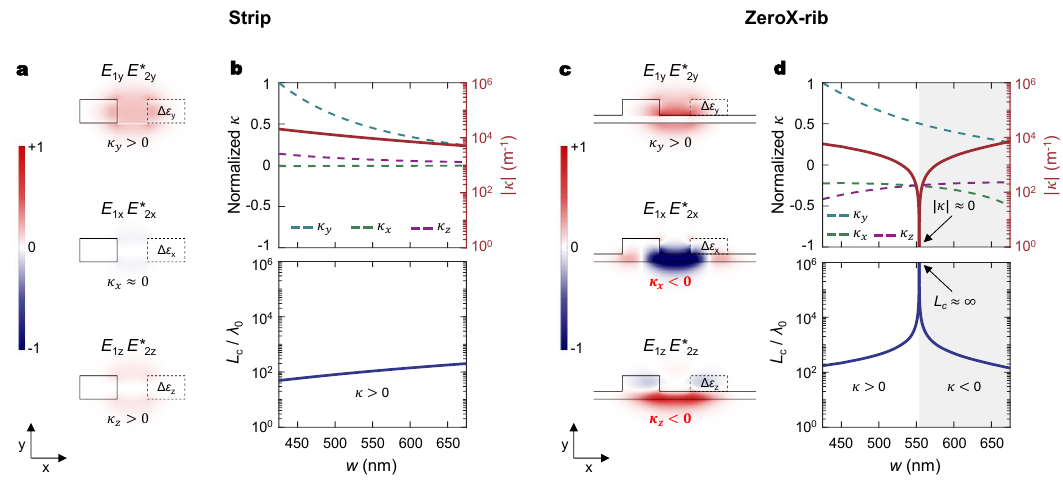}
	\caption{\textbf{Anisotropic coupling cancellation in zeroX-rib waveguides.}
        \textbf{a,} Conjugate products of the normalized electric-field components in coupled strip waveguides.
        Dashed lines denote the perturbation region used to calculate the coupling coefficients.
        \textbf{b,} Normalized coupling coefficients for strip waveguides as a function of core width $w$: $\kappa_y$ (cyan dashed), $\kappa_x$ (green dashed), and $\kappa_z$ (purple dashed).
        The total coupling coefficient $|\kappa|$ (red solid) and the corresponding normalized coupling length $L_\mathrm{c}/\lambda_0$ (blue solid) are also shown.
        \textbf{c, d,} Corresponding coupled-mode analysis for zeroX-rib waveguides: \textbf{c,} conjugate field-product distributions; \textbf{d,} coupling coefficients and the normalized coupling length.
        Negative $\kappa_x$ and $\kappa_z$ emerge owing to anisotropic mode hybridization.
        The shaded region in \textbf{d} denotes the regime where $\kappa<0$.
        The arrow marks the zero-crosstalk point where $|\kappa|\approx0$ and $L_\mathrm{c}/\lambda_0 \rightarrow \infty$.
       }
	\label{Fig3}
\end{figure*}

\noindent \textbf{Zero-crosstalk via anisotropic perturbation}\\
We next examine the physical origin of zero-crosstalk in slab-engineered rib waveguides. 
In conventional strip waveguides, all coupling components contribute with the same sign, yielding a finite positive coupling coefficient and unavoidable crosstalk.
In contrast, slab-engineered rib waveguides support sign-reversed coupling contributions that mutually cancel through anisotropic modal perturbations. 

Figures~\ref{Fig3}a,b show the coupled-mode analysis for strip waveguides. 
The coupling coefficients $\kappa_y$, $\kappa_x$, and $\kappa_z$ are calculated from the conjugate product of the corresponding electric-field components within the perturbation region (see Methods),  and their sum gives the total coupling coefficient $\kappa=\kappa_x+\kappa_y+\kappa_z$. 
The normalized coupling length is also evaluated \cite{Yariv2007photonics, Huang1994coupled},
\begin{equation}
\label{eq:coup_length}
\frac{L_\mathrm{c}}{\lambda_0} = \frac{\pi}{2|\kappa|\lambda_0}
\end{equation}
where larger $L_\mathrm{c}$ corresponds to lower crosstalk. 

For strip waveguides, the field distributions preserve their original symmetry, resulting in predominantly positive $\kappa_y$ and $\kappa_z$, while $\kappa_x$ remains negligible due to the weak $E_x$ field overlap.
As a result, all coupling contributions are constructive, yielding a finite positive $\kappa$ over the entire width range (Fig.~\ref{Fig3}b).
The corresponding coupling length is finite with $L_{\rm c}/\lambda_0\approx100$, implying that complete power transfer occurs with $\approx100$ wavelengths of propagation.

Figures~\ref{Fig3}c,d show the corresponding analysis for slab-engineered zeroX-rib waveguides.
Here, mode hybridization strongly redistributes the modal field profiles and reverses the sign of selected coupling components. 
Particularly, the anisotropic redistribution of $\operatorname{Re}(E_{x})$ gives a dominant negative overlap within the perturbation region, resulting in $\kappa_x<0$.
The vertically shifted bipolar distribution $\operatorname{Im}(E_{z})$ also exhibits negative $\kappa_z<0$.
In contrast, $\kappa_y$ remains positive due to the preserved polarity of the dominant TM field component.

Notably, these competing coupling contributions evolve differently as the core width $w$ increases. 
Enhanced modal confinement progressively reduces $\kappa_y$ and $\kappa_z$ through evanescent decay, whereas the magnitude of $\kappa_x$ increases due to stronger mode hybridization and enhanced $E_x$ field components (the TE$_{01}$ portion, see Supplementary Section~2).
This opposite evolution leads the total coupling coefficient toward zero, resulting in a non-trivial cancellation point, where $|\kappa|\approx0$ and $L_\mathrm{c}\rightarrow\infty$, as shown in Fig.~\ref{Fig3}d.

Complete cancellation occurs only within the weakly hybridized quasi-$\text{TM}_0$ regime, where the negative contributions from $\kappa_x$ and $\kappa_z$ become comparable to the positive contribution from $\kappa_y$.
For thinner slabs, the sign reversal does not occur, whereas excessively thick slabs produce overly large negative contributions for finite coupling (Supplementary Section~2).
These coupled-mode analyses are consistent with full-wave supermode simulations (Supplementary Section~3), validating the proposed coupling-cancellation mechanism.
\\

\begin{figure*}[!htbp]
	\includegraphics[width=\linewidth]{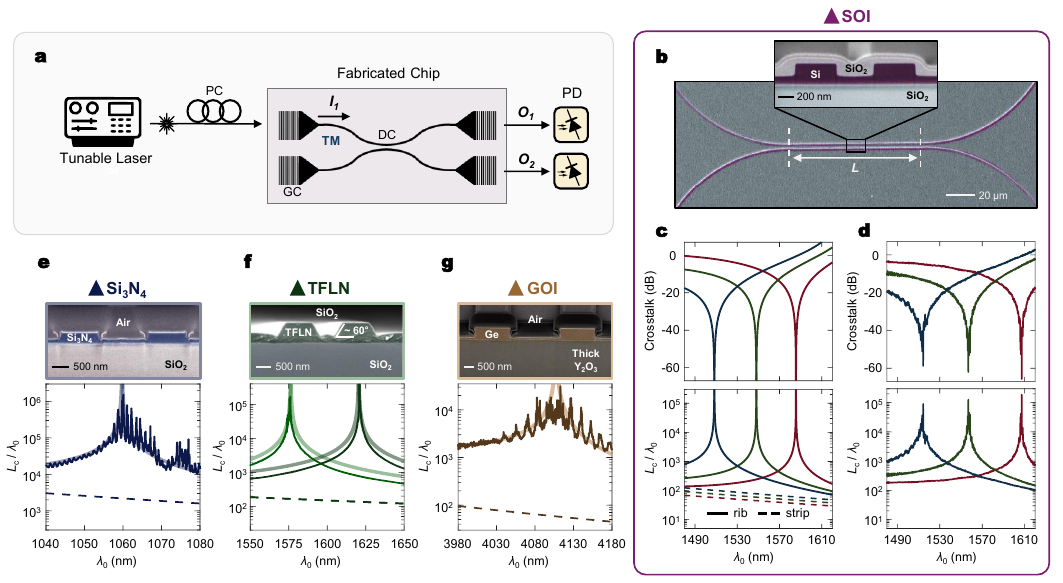}
	\caption{\textbf{Universal zeroX-rib photonic integration across material platforms and wavelength regimes.}
    \textbf{a,} Experimental configuration for crosstalk characterization.
    PC, polarization controller; GC, grating coupler; DC, directional coupler; PD, photodetector.
    $I_1$ is the input power after coupling through the GC; $O_1$ and $O_2$ are the output powers at the bar and cross ports, respectively.
    \textbf{b,} False-colored scanning electron microscope (SEM) image of the fabricated SOI zeroX-rib waveguides with the physical coupling length of $L=100~\mu\text{m}$.
    \textbf{c,d} Simulated (\textbf{c}) and measured (\textbf{d}) crosstalk spectra (top) and normalized coupling lengths $L_\mathrm{c}/\lambda_0$ (bottom) for strip (dashed) and zeroX-rib (solid) waveguides near 1550~nm. 
    Different colors represent different gap spacings: $g = 340$ (red), 370 (green), and 400~nm (navy). 
    \textbf{e--g,} Experimental validation of zeroX-rib on additional photonic platforms: \textbf{e,} $\mathrm{Si_3N_4}$ near $1060~\text{nm}$, \textbf{f,} Z-cut TFLN near $1550~\text{nm}$, and \textbf{g,} GOI (germanium core and yttrium oxide cladding) near $4200~\text{nm}$.
    Solid lines represent experimental results; faint lines denote the corresponding simulations; dashed lines indicate the simulated strip-waveguide counterpart.
    In all platforms, zeroX-rib waveguides achieve orders-of-magnitude longer coupling lengths than their strip counterparts.
    See Methods for additional parameters, and detailed experimental data are in Supplementary Section~4.
    }
	\label{Fig4}
\end{figure*}

\noindent \textbf{Universal zeroX-rib photonic integration}\\
We experimentally validate the universality of the proposed zeroX-rib concept across multiple photonic material platforms and wavelength regimes. 
Unlike platform-specific crosstalk-suppression approaches, the present mechanism emerges from slab-engineered mode hybridization inherent to rib geometries, making it broadly implementable across diverse photonic platforms.

We first validate the zeroX-rib mechanism on a SOI platform. 
Figure~\ref{Fig4}a shows the experimental configuration for crosstalk characterization, and Fig.~\ref{Fig4}b depicts scanning electron microscope (SEM) images of the fabricated device.
TM-polarized excitation is ensured using high-extinction-ratio grating couplers and polarization control, while stepped adiabatic tapers suppress higher-order mode excitation (Supplementary Section~4). 
The inter-waveguide crosstalk is extracted from the output power ratio between the bar and cross ports, yielding the normalized coupling length $L_\mathrm{c}/\lambda_0$ (see Methods).

Figure~\ref{Fig4}c shows the simulated crosstalk spectra and corresponding normalized coupling lengths for both strip (dashed) and zeroX-rib (solid) waveguides. 
Whereas strip waveguides exhibit strong coupling and short coupling lengths, the zeroX-rib waveguides exhibit pronounced crosstalk suppression and extremely long normalized coupling lengths. 
Notably, this enhancement is achieved at an identical waveguide pitch---directly translating to higher integration density. 
These predictions are experimentally verified in Fig.~\ref{Fig4}d, where the fabricated zeroX-rib waveguides exhibit crosstalk below $-60$~dB at the zeroX wavelength, limited by the photodetector noise floor. 
Different colors represent different gap spacings, enabling spectral tuning of the zeroX wavelength (here, $g=340$, 370, and 400~nm correspond to zeroX wavelength at $\lambda_0\approx1607$, 1557, and 1514~nm).
We also confirm consistent zeroX-rib operation in more deeply etched rib waveguides ($h_{\rm slab}=100$~nm), where the deeper etch broadens the operational bandwidth for crosstalk suppression (see Supplementary Section~4).
The achieved 20-dB crosstalk suppression bandwidth exceeds 60~nm, covering the bandwidth requirements of many PIC applications.
See Supplementary Section~5 for further parametric studies.

We further validate the universality of the zeroX-rib concept across distinct photonic material platforms and spectral regimes. 
Figures~\ref{Fig4}e--g show experimentally demonstrated zero-crosstalk operation across three additional platforms: $\mathrm{Si_3N_4}$ near $1060~\mathrm{nm}$, Z-cut TFLN near 1550~nm, and GOI near 4200~nm, respectively.
In all cases, the slab-engineered zeroX-rib structures exhibit strongly suppressed crosstalk and orders-of-magnitude longer coupling lengths than their strip counterparts, in good agreement with numerical simulations.
Crucially, these platforms (TFLN, GOI, $\mathrm{Si_3N_4}$) and wavelength regimes (near-infrared at 1060~nm and mid-infrared at 4200~nm) have lacked feasible crosstalk-suppression solutions due to their limited etchability, low index-contrast, and demanding fabrication requirements.
Our zeroX-rib approach thus opens crosstalk-free photonic integration to material platforms and wavelength regimes that have remained inaccessible to all previous methods, while remaining fully compatible with standard foundry processes.
Together, these results address the long-standing trade-off between signal fidelity and integration density across diverse material platforms and wavelength regimes.
This design principle remains effective for multi-waveguide systems (Supplementary Section~6), and can be further extended to heterogeneous photonic platforms \cite{Abel2019bto, Eltes2019bto, Churaev2023hetero,Niels2026hetero} with dissimilar slab and core materials (Supplementary Section~1), offering additional flexibility for advanced integrated photonic architectures.
\\

\noindent \textbf{\large Discussion and conclusion}\\
In summary, we have established slab-engineered zeroX-rib waveguides as a universal, scalable, and manufacturable route toward zero-crosstalk photonic integration. 
The mechanism---mode hybridization between TM$_0$ and TE$_1$ modes, controlled by slab thickness---induces anisotropic modal perturbations whose sign-reversed contributions to inter-waveguide coupling cancel within a weakly hybridized quasi-TM$_0$ regime. 
This cancellation principle is platform-independent: arising from intrinsic modal interactions rather than material-specific structures, it requires neither subwavelength patterning nor specialized fabrication, and therefore extends naturally across material platforms, wavelength regimes, and standard foundry processes. 
We experimentally validated this universality on SOI, $\mathrm{Si_3N_4}$, TFLN, and GOI platforms spanning wavelengths from the visible to the mid-infrared, achieving order-of-magnitude enhancements in coupling length and crosstalk suppression.

Beyond crosstalk suppression, slab-engineered mode hybridization establishes a broader design paradigm in which the slab thickness serves as a continuous control over inter-modal coupling---a degree of freedom that can be further exploited for polarization manipulation, dispersion engineering, and non-Hermitian phenomena in rib geometries. 
The zeroX-rib approach is directly applicable to large-scale photonic architectures---including dense waveguide arrays, photonic switching matrices, and integrated quantum circuits---where accumulated crosstalk has been a primary obstacle. 
Several extensions remain open: TE-polarized operation (with further parametric exploration using mode hybridization), curved and crossed geometries, and integration with active components (e.g., electro-optic modulators on TFLN/TFLT and BTO), each requiring dedicated mode-hybridization analysis under the same underlying principle.

By overcoming the long-standing trade-off between integration density and signal fidelity, the zeroX-rib concept establishes a scalable foundation for next-generation photonic systems---from co-packaged optics and photonic AI accelerators to large-scale quantum photonic technologies---where low-crosstalk, high-density integration across diverse material platforms is becoming ever more critical.

\section*{Methods}

\subsection*{\fontsize{10}{10}\selectfont Numerical simulations}
\noindent \textbf{Transverse electric polarization fraction:} A commercially available software package (Lumerical Mode Solutions) was used to calculate the TE polarization fraction in Fig.~\ref{Fig2}, given by:
\begin{equation}
\label{eq:te_polar}
\text{TE polarization fraction} = \frac{\int |E_{x}|^2 dx dy}{\int (|E_{x}|^2+|E_{y}|^2) dx dy},
\end{equation}
where $E_{x}$ and $E_{y}$ are the electric field components in the $x$ and $y$ axis, respectively. 
This parameter quantifies the relative field strength between $E_{x}$ and $E_{y}$, where a value of 0.5 signifies that their contributions become comparable.
This condition corresponds to the maximal mode hybridization between the TM$_{0}$ and TE$_{1}$ modes, where their relative contributions are nearly equal.
In all simulations, the $x$, $y$, and $z$ axes were set to the horizontal, vertical, and propagation directions, respectively.
\\\\
\noindent \textbf{Normalized coupling length:}
Lumerical Mode Solutions was used to calculate the normalized coupling length $L_\mathrm{c}/\lambda_0$ in Fig.~\ref{Fig4} and Supplementary Section~1.
We performed full-wave supermode simulations and calculated the normalized coupling length by \cite{Yariv2007photonics, Huang1994coupled}
\begin{equation}
\frac{L_\mathrm{c}}{\lambda_0} =\frac{1}{\Delta{n}}= \frac{1}{2|n_s-n_a|},
\label{eq:Lc_sim}
\end{equation}
where $n_{\text{s}}$ and $n_{\text{a}}$ are the effective indices of symmetric and antisymmetric modes, respectively, and $\lambda_0$ is the free-space wavelength.

\subsection*{\fontsize{10}{10}\selectfont Coupled-mode analysis}
The coupling coefficient $\kappa_i$ in Fig.~\ref{Fig3} was calculated using coupled-mode theory, following \cite{Yariv2007photonics, Huang1994coupled}:
\begin{equation}
\kappa_i= \frac{\omega\varepsilon_0}{4}\iint \Delta\varepsilon_i(x,y)
            {E_{1i}}(x,y) {E_{2i}^{*}}(x,y) dxdy
\label{eq:kappa}
\end{equation}
where $i=x,y,$ and $z$, representing the coupling coefficient along each axis.
${E_{1i}}$ and ${E_{2i}}$ correspond to the normalized electric field components of the $\mathrm{TM}_0$ modes (or quasi-$\mathrm{TM}_0$ modes) in each isolated waveguide, while $\Delta \varepsilon_i$ represents the dielectric contrast within the coupling region.
Note that, since both waveguides are geometrically identical, the field patterns ${E_{1i}}$ and ${E_{2i}}$ are equivalent, which means that ${E_{2i}}$ within $\Delta \varepsilon_2$ matches the field distribution shown in Fig.~\ref{Fig2}d-h.
The overall coupling coefficient $|\kappa|$ was obtained as the sum of all individual $\kappa_i$ components (i.e, $|\kappa|=|\kappa_x+\kappa_y+\kappa_z|$), and the corresponding normalized coupling length is then calculated by $L_\mathrm{c}/\lambda_0={\pi}/({2|\kappa|\lambda_0})$.
A more detailed method can be found in Ref.~\cite{Shin2025anisotropic}. \\

\subsection*{\fontsize{10}{10}\selectfont Device characterization}
\noindent \textbf{Crosstalk and coupling length characterization:} 
The experimental setup in Fig.~\ref{Fig4} was used to characterize the crosstalk and coupling length of the zeroX-rib waveguides. 
The tunable laser source, photodetector, and optical fibers were chosen to match the target measurement wavelength range; each component is described in detail below.
The crosstalk was quantified by measuring the output power ratio ($O_2/O_1$) between the bar ($O_1$) and cross ($O_2$) ports. 
From this ratio, the coupling length was extracted using the standard directional coupler relation \cite{Yariv2007photonics, Huang1994coupled}:
\begin{equation}
L_\mathrm{c} = \frac{\pi L}{2\tan^{-1}\!\sqrt{O_2/O_1}},
\label{eq:Lc_exp}
\end{equation}
where $L$ is the physical length of the directional coupler.
\\\\
\noindent \textbf{\boldmath$\mathrm{Si_3N_4}$ characterization (1010--1090~nm):} The photonic chips were characterized by a custom-built grating coupler setup. 
An angle-polished ($8^{\circ}$) eight-channel HI1060 fiber array was used to couple light in and out of the grating couplers. 
The fiber array was mounted on a five-axis stage with a high-precision adjuster of $50~\mu\text{m}$ per revolution in XYZ direction.
A Toptica tunable laser DLC-CTL 1060 was used as the source, and a Keysight 81635A optical power meter with InGaAs sensors was used as the output detector. 
The wavelength was swept from $1010~\text{nm}$ to $1090~\text{nm}$ with a step of $100~\text{pm}$. 
A polarization controller was used to control the polarization of the input laser light.
\\\\
\noindent \textbf{\boldmath SOI/TFLN characterization (1480--1650~nm):} 
The photonic chips were characterized using the same setup as described above, including the five-axis stage and polarization controller.
Depending on the target wavelength, an angle-polished ($8^{\circ}$) eight-channel SMF-28 fiber array was used for coupling, and a Keysight 81608A tunable laser and a Keysight N7744A optical power meter with InGaAs sensors were used as the source and the output detector, respectively.
The wavelength was swept from $1480~\text{nm}$ to $1650~\text{nm}$ with a step of $100~\text{pm}$. 
\\\\
\noindent \textbf{GOI characterization (3950--4200~nm):} The photonic chips were characterized by a custom-built mid-infrared (MIR) measurement setup, described in detail in Ref.~\cite{Lim2023low, Lim2024ultrasensitive, Shim2025room}. 
A mid-infrared quantum cascade laser (QCL) was used as the light source. 
The emitted beam was guided through free-space optical components, including mirrors, a half-wave plate, and an optical chopper, before being coupled into an $\mathrm{InF_3}$ fiber using a lens-coupled fiber holder. 
The optical signal was delivered to the device via a grating coupler. 
Precise fiber-to-chip alignment was achieved using an optical stage and a microscope. 
The output signal was collected and directed to a mercury cadmium telluride (MCT) photodetector as the output detector. 
To enhance the signal-to-noise ratio, the modulated optical signal was detected using a lock-in amplifier synchronized to the reference chopper frequency.

\subsection*{\fontsize{10}{10}\selectfont Device fabrication}
\noindent \textbf{SOI fabrication:} The photonic chips were fabricated on an SOI wafer with $300~\text{nm}$-thick Si and $2~\mu\text{m}$ $\mathrm{SiO_2}$ substrate.
The fabricated zeroX-rib waveguides have the following geometric parameters: for Fig.~\ref{Fig4}d, $h_\mathrm{slab} = 110~\mathrm{nm}$, $w = 500~\mathrm{nm}$, with $g = 340$ (red), 370 (green) and 400~nm (navy); for Supplementary Fig.~S9, $h_\mathrm{slab} = 100~\mathrm{nm}$, $w = 500~\mathrm{nm}$, with $g = 500~\mathrm{nm}$ (red) and $470~\mathrm{nm}$ (blue).
A JEOL JBX 8100FS electron beam lithography (EBL) system was used for fabrication. 
The operating conditions were 100~keV energy, 5~nA beam current, and $1000~\mu\text{m} \times 1000~\mu\text{m}$ field exposure. 
For sample preparation, Piranha and RCA1 cleaning was done, followed by dehydration bake on a $180~^\circ\text{C}$ hotplate for 30~min. 
HMDS was spin-coated at 5000~rpm and baked on a $95~^\circ\text{C}$ hotplate for 2~min. 
A negative tone resist (ma-N 2403, micro resist technology GmbH) was spin-coated at 3000~rpm and pre-exposure baked on a $95~^\circ\text{C}$ hotplate for 90~s. 
The base dose for exposure was $400~\mu\text{C/cm}^2$. The resist was developed using AZ MIF 300 for 50~s, followed by a deionized (DI) water rinse for 2~min. 
Then, nitrogen was blown in for air drying. Hard bake was done on a $110~^\circ\text{C}$ hotplate for 2~min to increase the stability of the resist pattern. 
The waveguide patterns were transferred into the Si layer by STS Multiplex ICP-RIE system with $\mathrm{CF_4/SF_6}$-based chemistry to an etch depth of 190 (Fig.~\ref{Fig4}d) and $200~\text{nm}$ (Supplementary Fig.~S9). 
The residual resist was fully stripped by Piranha and RCA1 cleaning. 
A $3~\mu\text{m}$-thick upper $\mathrm{SiO_2}$ cladding was then deposited by plasma-enhanced chemical vapor deposition (PECVD) using silane gas at $400~^\circ\text{C}$.
\\\\
\noindent \textbf{\boldmath$\mathrm{Si_3N_4}$ fabrication:} The photonic chips were fabricated on $300~\text{nm}$-thick $\mathrm{Si_3N_4}$ on $10~\mu\text{m}$ $\mathrm{SiO_2}$ substrate, using the same EBL procedure as for the SOI platform (see above).
The fabricated zeroX-rib waveguide in Fig.~\ref{Fig4}e has the following geometric parameters: $h_\mathrm{slab} = 110~\mathrm{nm}$, $w = 880~\mathrm{nm}$, and $g = 1080~\mathrm{nm}$.
The exposure dose was $375~\mu\text{C/cm}^2$, with development in AZ MIF 300 for 55~s. 
The $\mathrm{Si_3N_4}$ layer was etched using a NeoGEN-MAXISTM200L ICP-RIE system with $\mathrm{CF_4/CHF_3}$-based chemistry to an etch depth of $190~\text{nm}$. 
The residual resist was fully stripped by Piranha and RCA1 cleaning.
\\\\
\noindent \textbf{TFLN fabrication:} The photonic chips were fabricated on a $600~\text{nm}$-thick Z-cut TFLN on a 4.7~$\mu$m-thick $\mathrm{SiO_2}$ substrate.
The fabricated zeroX-rib waveguides in Fig.~\ref{Fig4}f have the following geometric parameters: $h_\mathrm{slab} = 250~\mathrm{nm}$, $w = 940~\mathrm{nm}$, with $g = 900~\mathrm{nm}$ (left) and $940~\mathrm{nm}$ (right).
A JEOL JBX-A9 electron beam lithography (EBL) system was used for fabrication.
The operating conditions were 100~keV energy, 5~nA beam current, and a $1000~\mu\text{m} \times 1000~\mu\text{m}$ exposure field.
A solvent rinse was done initially, followed by $\mathrm{O_2}$ plasma treatment at 350~W for 5~min for surface passivation before spin-coating. 
For waveguide and coupler patterning, a negative tone photoresist (ma-N 2405, micro resist technology GmbH) was spin-coated at 5000~rpm and pre-exposure baked on a $120~^\circ\text{C}$ hotplate for 150~s. 
To suppress charge effects during electron-beam exposure, a conductive e-spacer layer was additionally deposited on top of the resist stack. 
The exposure doses used were 300 and $350~\mu\text{C/cm}^2$. Following exposure, the resist was developed using AZ MIF 300 and rinsed in flowing deionized (DI) water for 5~min.
Then, nitrogen was blown in for air drying. 
The waveguide patterns were transferred into the TFLN by inductively coupled reactive ion etching (ICP-RIE) with Ar ions to an etch depth of $400~\text{nm}$. 
The residual resist was fully removed using 45\% potassium hydroxide (KOH) at $65~^\circ\text{C}$ for 45~min, and the chip was rinsed in flowing deionized (DI) water for 5~min.
A $2~\mu\text{m}$-thick upper $\mathrm{SiO_2}$ cladding was then deposited by plasma-enhanced chemical vapor deposition (PECVD).
Finally, an annealing process was carried out at $550~^\circ\text{C}$ for 3~hours in an $\mathrm{O_2}$ atmosphere.
\\\\
\noindent \textbf{GOI fabrication:} The photonic chips were fabricated on a $500~\text{nm}$-thick GOI platform with 3.2~$\mu$m-thick $\mathrm{Y_2O_3}$ substrate. 
The fabricated zeroX-rib waveguide in Fig.~\ref{Fig4}f has the following geometric parameters: $h_\mathrm{slab} = 75~\mathrm{nm}$, $w = 1270~\mathrm{nm}$, and $g = 1670~\mathrm{nm}$.
The GOI platform consists of a Si substrate and a Ge epitaxial wafer grown by metal-organic chemical vapor deposition (MOCVD). 
The Ge epitaxial stack comprised a $500~\text{nm}$-thick Ge core layer, a 5~nm $\mathrm{Si_{0.5}Ge_{0.5}}$ interlayer, and a 900~nm Ge strain-relaxed buffer (SRB) on a Si substrate. 
A 1.6~$\mu$m-thick $\mathrm{Y_2O_3}$ layer was deposited on both the Si wafer and the Ge epitaxial wafer via RF sputtering at $150~^\circ$C. 
Chemical mechanical polishing (CMP) was subsequently performed on both wafers to achieve sub-nanometer surface roughness. 
An additional 10~nm $\mathrm{Y_2O_3}$ layer was then deposited on each wafer, followed by $\mathrm{O_2}$ plasma surface activation using reactive ion etching (RIE) and direct wafer bonding. 
The GOI substrate was formed by selectively removing the sacrificial layers: the SiGe interlayer was etched using tetramethylammonium hydroxide (TMAH), while the Ge SRB layer was removed using an ammonium peroxide mixture (APM). 
Following substrate preparation, electron-beam lithography was employed to define photonic device patterns, including rib waveguides, directional couplers, and grating couplers, using a 200~nm-thick negative-tone resist (AR-N 7520). 
The Ge layer was then partially etched in multiple steps using inductively coupled plasma reactive ion etching (ICP-RIE) with $\mathrm{SF_6}$ (40~sccm) and $\mathrm{C_4F_8}$ (15~sccm) at 25~mTorr, RF power 50~W, and ICP power 600~W. 
The remaining Ge thickness was monitored after each etching step using a 3D laser profiler (VK-260K).

\begin{acknowledgments}
This work was supported by the National Research Foundation of Korea (RS-2026-25481462, RS-2023-NR119925, 2023R1A2C2002777) and the Institute of Information \& Communications Technology Planning \& Evaluation (RS-2026-25523959) grant funded by the Korea government (MSIT).\\
\end{acknowledgments}

\section*{Author contributions}
S.S.K. conceived and led the project. 
K.K. developed the idea and led the numerical simulations and coupled-mode analysis with contributions from Y.S. and M.L.
S.L., I.K., and H.Y.J. fabricated SOI/Si$_3$N$_4$, GOI, and TFLN devices, respectively, under the supervision of S.S.K., S.H.K., H.K., and H.J.J. 
K.K., Y.S., I.K., and J.S. characterized the devices.
K.K. and S.S.K. wrote the manuscript. 
All authors discussed the results and commented on the manuscript.

\bibliography{reference}

\end{document}